\documentclass[lettersize,journal]{IEEEtran}
\usepackage{amsmath,amsfonts}

\usepackage[linesnumbered,ruled,vlined]{algorithm2e}
\usepackage{algpseudocode}
\usepackage{array}
\usepackage[caption=false,font=normalsize,labelfont=sf,textfont=sf]{subfig}
\usepackage{textcomp}
\usepackage{stfloats}
\usepackage{url}
\usepackage{verbatim}
\usepackage{graphicx}
\usepackage{cite}
\begin{document}

\title{TrustFed: A Reliable Federated Learning Framework with Malicious-Attack Resistance}

\author{
	\IEEEauthorblockN{
		Hang Su, 
		Jianhong Zhou,
		Xianhua Niu, 
		Gang Feng 
	}
\thanks{H. Su, J. Zhou, and X. Niu are with the School of Computer and Software Engineering, Xihua University, Chengdu, China, G. Feng is with the National Key Lab on Communications, University of Electronic Science and
	Technology of China, Chengdu, China. J. Zhou is the corresponding author (e-mail:zhoujh@uestc.edu.cn).}
} 

\markboth{Journal of \LaTeX\ Class Files}%
{Shell \MakeLowercase{\textit{et al.}}: A Sample Article Using IEEEtran.cls for IEEE Journals}

\IEEEpubid{}

\maketitle

\begin{abstract}
As a key technology in 6G research, federated learning (FL) enables collaborative learning among multiple clients while ensuring individual data privacy. However, malicious attackers among the participating clients can intentionally tamper with the training data or the trained model, compromising the accuracy and trustworthiness of the system. To address this issue, in this paper we propose a hierarchical audit based FL (HiAudit-FL) framework, with aim to enhance the reliability and security of the learning process. The hierarchical audit process includes two stages, namely model-audit and parameter-audit. In the model-audit stage, a low-overhead audit method is employed to identify suspicious clients. Subsequently, in the parameter-audit stage, a resource-consuming method is used to detect all malicious clients with higher accuracy among the suspicious ones.  Specifically, we execute model audit method among partial clients for multiple rounds, which is modeled as a partial observation Markov decision process (POMDP) with aim to enhance the robustness and accountability of the decision-making in complex and uncertain environments. Meanwhile, we formulate the problem of identifying malicious attackers through a multi-round audit as an active sequential hypothesis testing problem and leverage a diffusion model-based AI-Enabled audit selection strategy (ASS) to decide which clients should be audited in each round. To accomplish efficient and effective audit selection, we design a  DRL-ASS algorithm by incorporating the ASS in a deep reinforcement learning (DRL) framework. Our simulation results demonstrate that HiAudit-FL can effectively identify and handle potential malicious users accurately, with small system overhead.
\end{abstract}


\section{Introduction}
\IEEEPARstart{W}{ith} the remarkable revolutionary progress in wireless communication technologies, the forthcoming wireless communication infrastructure will possess the capability of intelligence \cite{10081195}. By applying artificial intelligence (AI) technologies like machine learning (ML), wireless system can learn and predict various environmental characteristics (e.g., wireless resource requirements, traffic patterns, network composition, content requests, etc.) to achieve certain predefined targets. However, the effectiveness of these technologies heavily relies on the volume and quality of source data used for training AI models. In many scenarios, this source data exists in isolated silos, necessitating a unified approach to gather and process it for model training \cite{10.1145/3298981}. Nevertheless, concerns regarding data security and personal privacy protection have become more prominent in recent years, posing challenges to traditional data gathering and processing methods \cite{WOS:000552126200034}, \cite{Albrecht2016HowTG}. In response to these challenges, federated learning (FL) has emerged as a novel machine learning framework, aiming to address data silos while safeguarding data privacy \cite{9460016}. In the FL framework, multiple clients (e.g., mobile devices) and one or more central servers iteratively train an ML model to achieve desired global model accuracy \cite{9798758}. The clients, acting as data providers, perform local model training during each global iteration to achieve the target local model accuracy. On the other hand, the server aggregates the local models from clients to construct a global model with improved accuracy. Notably, only model-related parameters are shared, while the raw data remains securely stored on the client devices, ensuring user privacy. The proposition of federated learning addresses the privacy concerns arising from centralized data collection in traditional machine learning, paradigm, attracting significant attention from both academia and industry \cite{WOS:000500922700039}, \cite{WOS:000509905300001}.

In FL, the client training process is opaque to the upper application platform, which ensures the privacy protection for the clients. However, this non-transparency also implies that the client-side training process cannot be supervised, resulting in compromised model training quality. Meanwhile, as FL is a distributed learning framework, independent execution of the data training process by dispersed clients makes it susceptible to attacks from malicious clients \cite{9308910}. These attacks may involve injecting malicious data during local data collection or introducing poisoned parameters into local models being trained. Definitely, this kind of attacks will lead to degradation of global model accuracy and errors in decision-making based on the model. Moreover, as the aggregation results contain critical information beyond the data itself in the long-term FL process, a relatively concealed attack known as  free-riding attack appears in FL \cite{WOS:000659893802025}. Free-riders benefit from the collective resource (the global model) without contributing to the training process, pretending to participate and uploading dummy models without training local model \cite{1626427}. Although the free riding attack does not affect the iterative training process eventually converging to the desired goal \cite{MOTHUKURI2021619}, it poses risks of intellectual property loss and data privacy leakage during the FL process \cite{WOS:000659893802025}.

Existing research has indicated that external audit methods can effectively defend against malicious attacks initiated by the clients \cite{9308910}. These methods involve reviewing local models uploaded by clients before each round of model aggregation to identify and reject malicious models \cite{WOS:000452649400012}. Nonetheless, current audit methods face a formidable challenge in maintaining a high level of audit accuracy with reasonable associated audit resource overhead.

To accurately identify malicious clients without excessive computational overhead, in this paper we propose to embed a hierarchical audit scheme into the legacy FL framework, namely HiAudit-FL. The proposed audit scheme combines a low-overhead model audit method with low audit accuracy, and a high-overhead parameter audit method with high audit accuracy. By leveraging both methods, the audit scheme can detect the malicious nodes efficiently for enhancing the reliability of the FL process. The main contributions of this paper is be summarized  as follows:

\begin{itemize}
	\item [1)]
	We propose a reliable federated learning framework by introducing a hierarchical audit mechanism, which significantly elevates the resilience against adversarial attacks in traditional federated learning paradigms.
	\item [2)]
	Within the proposed framework, we formulate the problem of identifying malicious attackers as an active sequential hypothesis testing problem and employ a multi-round partial clients audit method. We derive the expressions for posterior probabilities of hypothesizes, which is a critical step in quantifying and enhancing the accuracy of the audit process.
	\item [3)]
	We present the AI-Enabled client selection strategy (ASS), empowered by diffusion models, to decide which clients should be audited in each round in the face of environmental uncertainty. We apply our proposed ASS to deep reinforcement learning (DRL) algorithm, which achieves efficient and optimized clients selection, therevy reducing audit overhead effectively while maintaining the required audit accuracy.
\end{itemize}

The rest of the article is organized as follows. Section II provides a review of  related work. The proposed HiAudit FL framework is elaborated in Section III. Section IV and Section V formulates the audit selection problem and develops a diffusion model-based DRL-ASS algorithm to solve the problem, respectively.  Section IV formulates the audit selection problem and Section V presents a diffusion model-based DRL-ASS algorithm to solve the problem. Finally, the paper is concluded in  Section VII.

\section{Related Work}
In centralized machine learning systems, attacks usually originate from external  attackers unrelated to the training process, such as malicious attackers outside the server launching attacks on the data center to steal data. The FL system adopts a distributed learning strategy, which increases the difficulty for attackers outside the system to steal data information. However, the distributed training framework brings new attacking risk, which is initiated by participating clients. The reason is that the raw data used for model training is hold by individual clients, instead of uploading to the central sever. However, the attacks initiated by clients within the system are still the primary source of threats to the security of the FL process \cite{9415623}. Next, we elaborate the related works from the following two aspects: the  threats initiated by clients to the FL process and  recent solutions to the corresponding attacks.

\subsection{Main attacks initiated by clients to the FL process}
As FL has been proposed and applied in many fields, such as finance, business, and others, the attacks and intrusions for FL systems have gradually become diverse. However, all attacks initiated by  FL clients can generally be  categorized into the following types:
\begin{itemize}
	\item [1)]
	Privacy inference attack. This attack aims to infer some specific information about the original training data from the FL learning model \cite{10044183}.
	\item [2)]
	Data poisoning attack. Data poisoning attack refers to the attacker injecting malicious samples into its local data, aiming to damage the performance of the global model by affecting the local training process. 
	\item [3)]
	Model poisoning attack. A model poisoning attacker may guide the global model to learn wrong knowledge by submitting false local model updates, causing the final model to produce incorrect predictions or poor prediction performance. For example, the authors in \cite{bagdasaryan2020backdoor} did some experiment on how backdoor attacks are implemented.
	\item [4)]		
	Free-rider attack. Free-rider attack in FL was revealed by the authors of \cite{lin2019free}. Specifically, this attack means that some malicious parties try to obtain model updates contributed by other parties without actually participating in calculating and uploading model updates.  In other words, the attackers aim to capture the final global model while avoiding the sacrifice of one's own computing resources and data privacy.
\end{itemize}

\subsection{Main solutions to the corresponding attacks}
Existing typical defense mechanisms  against client-initiated attacks belong  reactive defense \cite{MOTHUKURI2021619}, i.e., actions taken in response to identified attacks. 

To defend against the privacy inference attack effectively, the differential privacy (DP) technology is used  to protect privacy by adding noise to personal sensitive attributes \cite{dwork2008differential}. Meanwhile, in \cite{10091486} differential privacy is applied by adding Gaussian noise to the gradients of local model to preserve training data and hidden personal information against external threats while the convergence property is guaranteed. To defend against the data poisoning attacks, the most critical technique is data cleaning. For example,  the research work in \cite{li2021sample} designs an efficient data quality evaluation method based on influence function to evaluate the negative influence of error in local training data on the model. Thus, the performance of federated learning system can be improved by eliminating the data with error.

To defend against the model poisoning attacks and free-rider attacks initiated by internal attackers, abnormal models can be identified through the abnormal audit mechanism \cite{MOTHUKURI2021619}. The authors of \cite{an2015variational} proposed to use  variational autoencoders for spectral anomaly auditing to identify model poisoning attacks from malicious clients. Lin  et al. revealed the existence of free-rider attacks in FL and proposed a new high-dimensional detection method called STD-DAGMM, which distinguishes existing free-riders through the energy value of each model \cite{lin2019free}. However, these abnormal auditing solutions still cannot accurately identify all potential malicious clients in scenarios with multiple malicious clients, and these solutions also have high audit errors, which means misjudgment of real or malicious clients. If an identification and elimination strategy is adopted,  it will face the challenge of honest clients being mistakenly canceled from training opportunities. In order to further ensure the reliability of  FL training process, Wang et al. \cite{9839627} proposed a parameter audit method for the client training process. Under the proposed MAM mechanism, the audit process is added to the federated learning system. All clients are required to save and upload all data parameters relevant to the training process, including the original data set of training, parameters generated by the gradient descent process, during each round of training. The audit platform will audit these parameters and identify malicious clients. This method can identify model poisoning attacks and free-rider attacks in FL systems with high accuracy. However, this parameter auditing strategy  incurs  massive additional system overhead due to auditing the parameters of all participants in each training round.

In summary,  aforementioned existing audit strategies  against client-level attacks in FL still cannot guarantee higher accuracy with  low computing overhead. This motivates us  to propose a hierarchical audit mechanism to reduce system computing overhead while ensuring audit accuracy.

\section{HiAudit-FL Framework}
In this section, we describe the proposed HiAudit-FL system  and then analyze the overhead incurred  by the audit process. 
\subsection{The HiAudit-FL Framework}
\begin{figure}[htbp]
	\centerline{\includegraphics[width=1\linewidth]{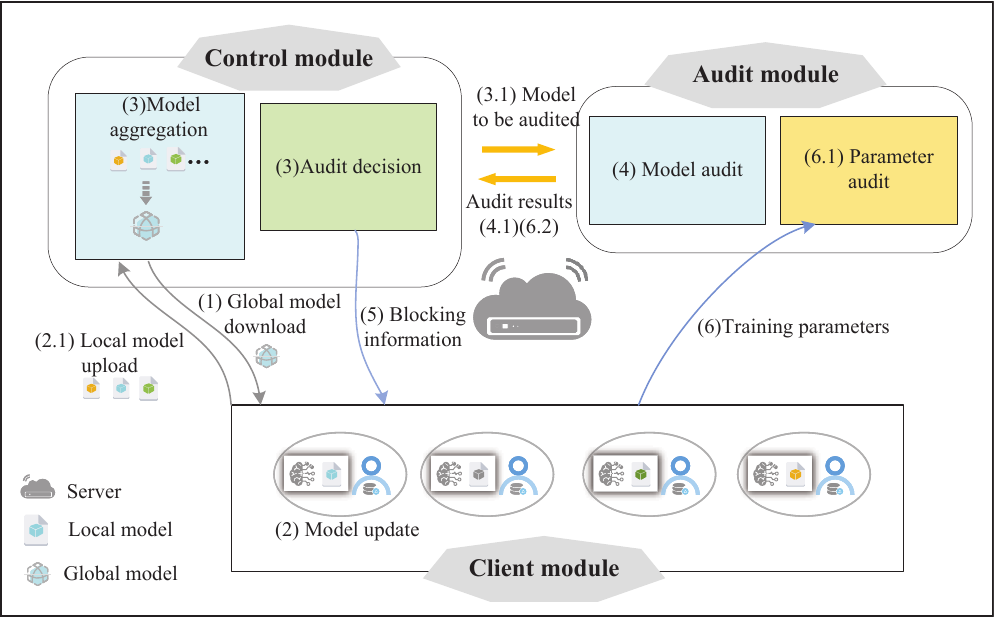}}
	\caption{HiAudit-FL framework}
	\label{fig1}
\end{figure}

The proposed hierarchical audit-based FL framework, called HiAudit-FL, is depicted in Fig.~\ref{fig1} This framework consists of three logical modules: client, control, and audit modules. The client and control modules largely retain the structure of the legacy FL framework, while  our designed audit module is incorporated into the legacy framework , addressing the issues arising from the non-transparency of the client's local training process, which could result in malicious data injection or harmful local model uploads. The client module handles local training, uploads  trained local models, and receives global models from the server.  The control module receives the trained local models, performs parameters aggregation, receives audit results from the audit module, and facilitates collaboration between the client module and the audit module. Obviously, the latter two functions of the control module  legacy FL framework. The audit module performs  audit-related operations and provides  audit results to control module for decision-making. Both the control module and audit module functions can be executed by the parameters aggregation servers, eliminating the need for new components in comparison to the legacy FL system.
The specific learning process under the HiAudit-FL system can be summarized as follows.
\begin{itemize}
	\item [1)]
	The control module sends the global model to all clients.
	\item [2)]
	The clients perform model training and upload their local models to the control module.
	\item [3)]
	The control module aggregates models, and selects some clients based on the proposed AI-Enabled client selection strategy (ASS), which will be elaborated in detain in the section V.
	\item [4)]		
	The audit module performs model audits on the selected clients in step 3). 
	\item [5)]
	Steps 1)-4) are repeated for several rounds until the model audit results meet the expected threshold. 
	\item [6)]
	The control module sends control information to the clients, instructing them to stop ongoing local training process and initiate a parameter audit request for some clients.
	\item [7)]
	The audit module performs the parameter audit process on these clients and share the audit result.
	\item [8)]
	Based on the parameter audit result, the control module decides whether to eliminate malicious clients and updates the global model to a clean model. 
	\item [9)]
	Steps 1)-7) are repeated until the global model reaches the specified accuracy or the expected training rounds.
\end{itemize}

\subsection{The Hierarchical Audit Scheme}
\begin{figure*}[htbp]
	\centerline{\includegraphics[width=\linewidth]{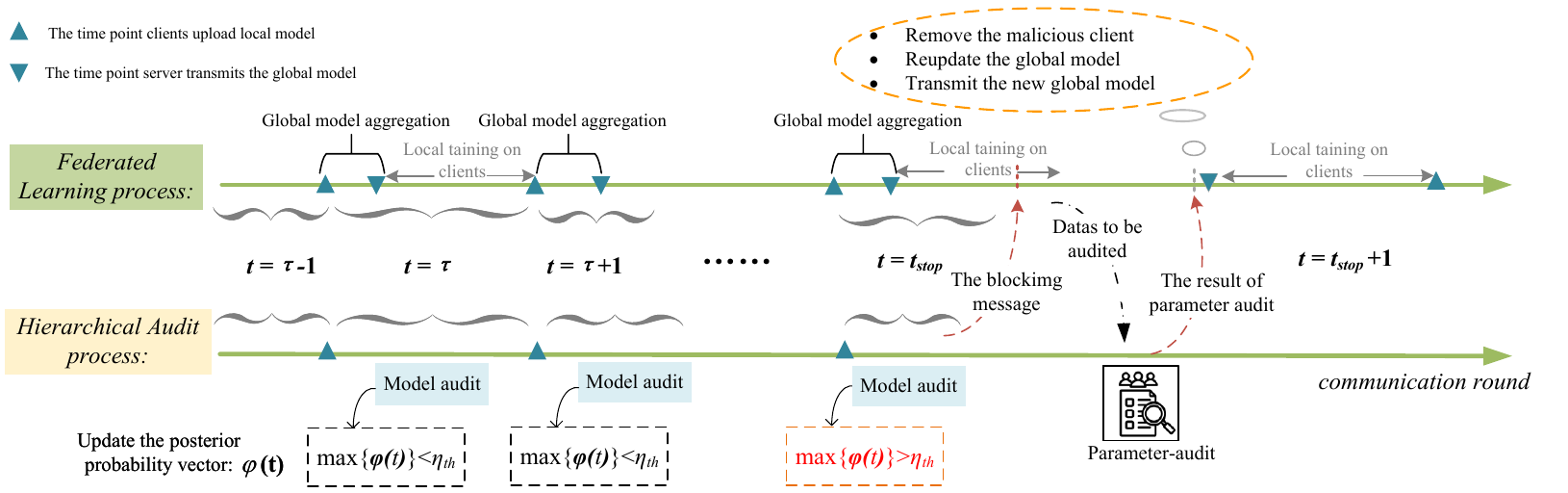}}
	\caption{Hierarchical audit scheme execution process}
	\label{fig2}
\end{figure*}

Next we present a detailed illustration of the proposed hierarchical audit (HiAudit) scheme, implemented within the audit module as depicted in Fig.~\ref{fig2}. The scheme comprises a model audit method and a parameter audit method. And the procedure is synchronized with the FL training process, with global learning round and audit round occurring simultaneously. We denote the $t$-th global learning or audit round with the symbol “$t$” as depicted in Fig. 2. 

After each round of receiving the local models uploaded by the clients, the server selects some of these models based the proposed ASS strategy to undergo the low-level model audit. This audit process generates a posterior probability vector as the audit result, which is recorded in the server log. With the increase in the number of audit rounds, the posterior probability vector will be updated until a value in the vector exceeds the specified threshold, indicating potentially malicious clients. The derivation of the posterior probability vector will be elaborated in detain in the subsequent section. When this threshold is reached at the end of the audit round ($t_{stop}$  as shown in Fig.~\ref{fig2}), the server announces to block the FL training process and coordinates a high-level parameter audit for these highly likely malicious clients to address the uncertainty associated with the probabilities and prevent honest clients from being wrongly eliminated. Upon completion of the parameter audit process, the malicious clients are identified and disqualified from further participation in the FL training. The sever updates the global model accordingly, and the FL training process continues among the remaining clients.

The improved accuracy of the posterior probability obtained through model audit allows us to identify potential malicious clients more accurately, thereby reducing the overhead caused by parameter audit. To achieve a more precise posterior probability vector with less model-audit overhead, we employ a multi-round partial clients audit method instead of auditing all clients multiple times. Specifically, during the model audit stage, we select local models uploaded by some clients to execute the model audit and output a probability vector at the end of each audit round. The posterior probability is updated by combining the probability vector obtained in the current round and the probability vectors obtained in all previous rounds. Based on the updated posterior probability, some clients are selected for the next round of model audit to enhance the audit accuracy efficiently. Obviously, how to determine the clients that will be selected to perform model audit in each round to achieve higher posterior probability accuracy with fewer audit rounds and smaller system overhead is a challenge. Thus, to obtain an optimized client selection strategy to reduce the audit overhead while ensuring audit accuracy, we model the audit related overhead and formulate the partial client selection problem in the next section. 

\section{Overhead Analysis and Problem Formulation}
\subsection{System Overhead Analysis}
We adopt a quasi-static state model \cite{8963610} for the system, which assumes that no clients actively join or leave during FL process. And we assume that if a client sends the malicious model at the first global iteration round, it will remain malicious throughout all rounds during the FL process. Since HiAudit-FL preserves the local training and model uploading process for honest clients as in traditional FL framework, we primarily focus on the additional system overhead caused by malicious clients, excluding the overhead of the FL process itself. Thus, the system overhead includes model-audit overhead, parameter-audit overhead and the malicious clients retention overhead.

\paragraph{Model audit Overhead}The audit overhead pertains to the computational resources utilized for executing the model audit algorithm. We denote the computation overhead used to audit a model as $\nu$. As all uploaded local models have the same size, the model-audit overhead depends on the number of selected clients. Thus, the model-audit overhead $\mathcal{C}_{model}$ can be expressed as:
\begin{equation}
	\mathcal{C}_{model}=\sum_{t=1}^{t=t_{stop}}\nu*\chi_{t},\label{eq1}
\end{equation}

\paragraph{Parameter audit overhead}Likewise, parameter audit necessitates clients to upload all training process parameters and reproduce their local iterative process for verifying training authenticity. This verification process incurs computational   overhead. Let $\varrho$ denote the computation overhead required to compute one data sample, and $d_i$ represent the amount of data declared by client $i$ for auditing. Therefore, the aggregate computational overhead for performing parameteraudit can be given by  
$\mathcal{C}_{para}$,
\begin{equation}
	\mathcal{C}_{para}=\sum_{i\in \boldsymbol{Z_{t_{stop}}}}\varrho*d_{i},\label{eq2}
\end{equation}
where $\boldsymbol{Z_{t_{stop}}}$ represents the set of clients selected undergo parameter audit at the audit round $t=t_{stop}$.

\paragraph{Malicious client retention overhead}In federated learning, malicious clients staying in the system may cause  two parts of overhead: invalid reward overhead and intellectual property loss overhead. The invalid reward overhead comes from incentives issued by the server to malicious clients. We assume that the system incentivizes client participation in the learning process by issuing rewards. However, the local models uploaded by malicious clients do not genuinely contribute to the improvement of the global model, and may even  have negative impact. Therefore, the reward issued to the malicious client by the server is invalid. We assume that the system assigns an average reward $\rho$ to each client in every round of global training, and the invalid reward issued by the system is correlated with the number of rounds the malicious client stays in the system. Besides, the global model leaked to malicious clients will result in loss of intellectual property, which is related to the number of malicious clients that remain in the system at the end of federated learning. Consequently, the malicious client retention overhead $\mathcal{C}_{mal}$ can be expressed as follows,
\begin{equation}
	\mathcal{C}_{mal}=\rho\sum_{t=1}^{t=T}n_t+n_T\mathcal{K}  ,\label{eq3}
\end{equation}
where $n_t$ represent the number of malicious clients remaining in the system at the round $t$, $T$ denote the maximum global iteration rounds of FL. And $\mathcal{K}$ denote the intellectual property loss caused by global model leakage.

\subsection{Client Selection Problem formulation}
We consider that $N$ learning clients and one server participate in the federated learning. The client $i$, ($1\leq {i} \leq {N}$) has a private training dataset $\boldsymbol{D}_i=\{(b_{i,1},c_{i,1}),(b_{i,2},c_{i,2}),...\}$, where $(b_{i,1},c_{i,1})$ represents the feature vector and corresponding label of the training samples for the client $i$, respectively. After receiving the initialized global model sent by the server, the eligible client $i$ begins to perform the local training process. The local training goal is to minimize the loss function and obtain a model $\boldsymbol\omega_i(t)$ as 
\begin{equation}
	f_n(\boldsymbol\omega_i(t))=\frac{1}{|\boldsymbol{D}_i|}\sum_{k\in \boldsymbol{D}_i}L_i(b_{i,k},c_{i,k};\boldsymbol\omega_i(t)),\label{eq4}
\end{equation}
where $L(\bullet)$ represents the loss function for a single data sample. After finishing the local training process, all the local training models will be transmitted to the server, where  specific parameter aggregation algorithm can be used to aggregate all the local training model  and update the global model, such as FedAvg \cite{pmlr-v54-mcmahan17a}, FedProx \cite{MLSYS2020_1f5fe839} and SCAFFOLD \cite{WOS:000683178505024}. For  ease of expression, we adopt the commonly used FedAvg algorithm to realize the parameters aggregation, which is expressed as:
\begin{equation}
	\boldsymbol\omega(t+1)=\sum_{i=1}^{N}\frac{|\boldsymbol{D}_i|}{|\boldsymbol{D}|}\boldsymbol\omega_i(t),\label{eq5}
\end{equation}
where $|\boldsymbol{D}|=\sum_{i=1}^{N}|\boldsymbol{D}_i|$.
Due to the limitation of model-audit algorithm, the model- audit state cannot accurately reflect the actual state of the clients, which means that audit error exists. Without loss of generality, we regard the audit error of the model-audit algorithm as a random event and  denote the probability of error event as $q$, ($0<q<1$). We use vector $\boldsymbol\mu(t)=[\mu_1 (t),\mu_2 (t),…,\mu_N(t)]\in\left \{ 0,1 \right \}^N$ to represent the actual state of the clients at the round $t$, where $\mu_i(t)=0$ means the client $i$ is honest, and $\mu_i(t)=1$ means that client $i$ is malicious. At the same time, we stipulate that binary variable $\lambda_i(t)$ represents the audit state of the client $i$ at the round $t$. Thus, we have:
\begin{equation}
	\lambda_{i}(t)= \left\{\begin{array}{rcl}
		1-\mu_i(t) & \mbox{with probility $q$} \\
		\mu_i(t) & \mbox{with probility $1-q$}
	\end{array}\right..\label{eq6}
\end{equation}
Obviously, the audit accuracy $\varepsilon_i(t+k)$ for the client i at audit round t can be expressed by the error probability as $1-q$. Theoretically, if we audit all clients in all rounds, after $k$ rounds model audit, the audit accuracy $\varepsilon_i(t+k)$ could be improved as ${1-q}^k$. However, the complete audit for all clients in every round may incur significant overhead. We thus consider that only a part of the clients will be chosen to be audited. Obviously, too many clients selected in each round may increase the audit accuracy to a great extent, but at the same time it will incur substantial  overhead. Conversely, if the number of clients selected in each round is too small, although it can save overhead, it may lead to a decrease in audit accuracy. Thus, an optimal client selection strategy should help us adaptively select the clients for model-audit at the beginning of each audit round, and effectively reduce system overhead while ensuring the audit accuracy. 

In summary, the main goal for the HiAudit scheme is to implement model-audit and parameter-audit to identify  and eliminate all malicious clients accurately. Thus, the server needs to make the following decisions at each audit round:
\begin{itemize}
	\item [(1)]
	Whether the parameter audit should be applied and the blocking operation should be performed?
	\item [(2)]
	If the parameter audit operation should not be performed, which clients should be selected to continue the model-audit process?
\end{itemize}
There are two goals in the decision process for the server: to specify the blocking rule and to obtain the clients selection strategy. We stipulate that the control module sends a blocking message and parameter-audit requests to the client module to stop the local training when the largest audit accuracy result among all N audit accuracy results reach the required audit accuracy $\varepsilon_{th}$ for the first time. Thus, the blocking rules could be expressed as 
\begin{equation}
	\max\{\varepsilon_i(t_{stop}) \}>\varepsilon_{th}, i\in\{1,N\}.\label{eq7}
\end{equation}
Thus, we formulate the client selection problem as 
\begin{equation}
	\begin{aligned}
		\min_{\{\Phi_t \}_{t=1}^{t=t_{stop}}} \mathcal{C}_{model}+\mathcal{C}_{para}+\mathcal{C}_{mal} \\
		s.t.  \max\{\varepsilon_i(t_{stop}) \}>\varepsilon_{th}, i\in\{1,N\},\label{eq8}
	\end{aligned}
\end{equation}
where $\Phi_t$ is the subset of the $N$ clients at round $t$, and the $t_{stop}$ means that after the round $t=t_{stop}$, the largest audit accuracy result among all $N$ audit accuracy results reaches the required audit accuracy $\varepsilon_{th}$ for the first time. The audit accuracy $\varepsilon_i(t)$, which we need to solve the problem can be obtained only if both the audit state $\lambda_i(t)$ and the actual state $\mu_i(t)$ are known. Unfortunately, the $\mu_i(t)$ cannot be known in real scenarios, which bottlenecks the problem solving. 

As mentioned before, the main objective of model audit in the Hierarchical Audit scheme is to obtain the malicious probability of all clients so as to guide the implementation of subsequent parameter audit. Thus, we can formulate the audit accuracy of the model-audit as a posterior probability whether the client is malicious or not. In this case, we can effectively use the condition that the actual state $\mu_i(t)$ can be inferred from the audit state $\lambda_i(t)$, which is representative for an active sequential hypothesis testing problem \cite{WOS:000461021200104}. Next we define  the model-audit accuracy as a posterior probability and formulate it as an active sequential hypothesis testing problem. 

Specifically, in a federated learning scenario with N clients, there are $h=2^N$ kinds of hypotheses. $\mathbf{H}=\left\{H_0,H_1,...,H_{h-1}\right\}$ is a set of hypotheses, and each hypothesis maps to a possible state combination of all the clients in the system. For example, we assume $N=5$, and $H_0$ (the 5-bit binary representation of subscript is “00000”) indicates that there is no malicious client in the current system, $H_1$ ("00001”) denotes that client 5 is malicious, $H_3$ (“00011”) indicates that client 4 and client 5 are malicious. 

After mapping the clients state to the hypothesis, the blocking rule need to be re-defined. We introduce an event vector $\mathbf{E}(t)=\left\{E_0(t),...,E_{h-1}(t)\right\}$, where $E_j(t), (0\le j\le h-1)$, represents the event that the hypothesis $H_i$ is true at audit round $t$. 

And the belief is the posterior probability of an event $E_j(t)$ conditioned on the audit result $\boldsymbol\lambda(t)$ available for the decision:
\begin{equation} 
	\mathbb{P}\{E_j (t)\}=\mathbb{P}\Big\{\boldsymbol\mu(t)=Bin_N\{j\} \Big| \{\boldsymbol\lambda_  {\Phi_{\tau}}   (\tau) \}_{\tau=1}^{\tau=t} \Big\},\label{eq9}
\end{equation}
where $Bin_N\{j\}$ is  an $N$-bit binary conversion function, it can convert a decimal number $j$ to an binary vector which size is $N$, we assume $N=5$, 
\begin{equation}
	\begin{aligned}
		Bin_5(0)=[0,0,0,0,0]\\
		Bin_5(1)=[0,0,0,0,1]\\
		Bin_5(2)=[0,0,0,1,0]\\
		...  \label{eq10}
	\end{aligned}
\end{equation}
Based the belief, we stipulate that the new blocking rule as follows:
\begin{equation}
	\resizebox{.98\hsize}{!}{$
		t_{stop}=\min \{ t>0: \max\{\mathbb{P}\{E_0 (t)\},…,\mathbb{P}\{E_{h-1}(t)\}\}>\eta_{th} \}, $}
	\label{eq11}
\end{equation}
where $\eta_{th}\in(0,1)$ is a predefined upper threshold on the belief, which is related to the expected result accuracy of multi-round partial audit.

For the convenience of calculation, we denote the posterior belief vector as $\boldsymbol\varphi(t)=[\varphi_0(t),...\varphi_j(t),…,\varphi_{h-1}(t)]$, where the index $j$ of vector indicates the hypothesis  $H_j$ and the value $\varphi_j(t)$ indicates the posterior belief that $H_j$ is true. Thus, the $\varphi_j(t)$ can be given by:
\begin{equation}
	\varphi_j(t)=\mathbb{P}\Big\{ \boldsymbol\mu(t)=Bin_N\{j\} \Big| \{\boldsymbol\lambda _{\Phi _\tau} (\tau)\}_{\tau=1}^{\tau=t}\Big\}.\label{eq12}
\end{equation}
According to Bayes' rule, $\varphi_j(t)$ can be expressed as:
\begin{equation}
	\resizebox{.98\hsize}{!}{$
		\varphi _j(t)=\frac{\mathbb{P} \big\{\{\boldsymbol\lambda_  {\boldsymbol\Phi_{\tau}}   (\tau) \}_{\tau=1}^{\tau=t} \big|Bin_N\{j\}\big\}*\mathbb{P}\big\{\boldsymbol\mu(t)=Bin_N\{j\} \big\} }{\sum_{k=0}^{h-1}\mathbb{P}\big\{\{\boldsymbol\lambda_  {\boldsymbol\Phi_{\tau}}   (\tau) \}_{\tau=1}^{\tau=t}\big|Bin_N\{k\}\big\}*\mathbb{P}\big\{\boldsymbol\mu(t)=Bin_N\{k\} \big\}}, 
		$}\label{eq13}
\end{equation}
We denote the actual state transition actual state transition probability at audit round $t$ as:
\begin{equation}
	\boldsymbol{P}_{kj}[t]=\mathbb{P}\Big\{\boldsymbol\mu(t)=Bin_N\{j\} \Big| \boldsymbol\mu(t-1)=Bin_N\{k\}\Big\},\label{eq14}
\end{equation}
and because we assume that the malicious clients will obtain the malicious state in quasi-static state, so the state transition probability can be denoted as:
\begin{equation}
	\boldsymbol{P}_{kj}[t]= \left\{\begin{array}{rcl}
		1 & \mbox{if j=k}\\
		0 &\mbox{otherwise}
	\end{array}\right. ,\label{eq15}
\end{equation}
and we further simplify (13) as follows:
\begin{equation}
	\begin{small}{
			\begin{split}
				&\mathbb{P}\Big\{\{\boldsymbol\lambda_  {\Phi_{\tau}}   (\tau) \}_{\tau=1}^{\tau=t}\big|Bin_N\{j\}\Big\}*\mathbb{P}\Big\{\boldsymbol\mu(t)
				=Bin_N\{j\} \Big\}\\
				&=\sum_{k=0}^{h-1}\bigg[\mathbb{P}\Big\{\boldsymbol\lambda_  {\Phi_{t}} (t) \big| \boldsymbol\mu(t)=Bin_N\{j\} \Big\}\\
				&*\boldsymbol{P}_{kj} [t-1]
				*\mathbb{P}\Big\{\boldsymbol\mu(t-1)=Bin_N\{k\} \big| \{\boldsymbol\lambda_  {\Phi_{\tau}}   (\tau) \}_{\tau=1}^{\tau=t-1} 	\Big\}	\bigg]\\
				&=\mathbb{P}\Big\{\boldsymbol\lambda_  {\Phi_{t}} (t) \big| \boldsymbol\mu(t)=Bin_N\{j\} \Big\}*\sum_{k=0}^{h-1}\Big[	\boldsymbol{P}_{kj}[t-1]*\varphi_k(t-1)\Big]\\
				&=\mathbb{P}\Big\{\boldsymbol\lambda_  {\Phi_{t}} (t) \big| \boldsymbol\mu(t)=Bin_N\{j\} \Big\}	\boldsymbol{P}_{jj}[t-1]*\varphi_j(t-1),\label{eq16}
			\end{split}
	}\end{small}
\end{equation}
where $\mathbb{P}\big\{\boldsymbol\lambda_  {\Phi_{t}} (t) \big| \boldsymbol\mu(t)=Bin_N\{j\} \big\}$ is expressed as:
\begin{equation}
	\begin{aligned}
		\vartheta_j(t)&= \mathbb{P}\Big\{\boldsymbol\lambda_  {\Phi_t}(t) \big| \boldsymbol\mu(t)=Bin_N\{j\} \Big\}\\
		&=\prod_{a\in \Phi_t} q^{\boldsymbol{1}\big\{ \boldsymbol\lambda_a(t)\neq Bin_N\{j\}_a\big\}}(1-q)^{\boldsymbol{1}\big\{ \boldsymbol\lambda_a(t)= Bin_N\{j\}_a\big\}},\label{eq17}
	\end{aligned}
\end{equation}
here $\boldsymbol{1}\{.\}$ is the indicator function. Finally we can re-express the $\varphi_j(t)$ as follows:
\begin{equation}
	\varphi_j(t)=\frac{\varphi_j(t-1)*	\vartheta_j(t)}{\sum_{k=0}^{h-1}\varphi_k(t-1)*	\vartheta_k(t)}.\label{eq18}
\end{equation}
Thus, the new blocking rule can be re-written as
\begin{equation}
	t_{stop}={t>0:\max \big\{ \boldsymbol\varphi(t)\big\}> \eta_{th}}\label{eq19}
\end{equation}
which means that the posterior probability of a hypothesis is true is greater than the threshold $\eta_{th}$. The client selection problem can be re-modeled as 
\begin{equation}
	\begin{aligned}
		\min_{\{\Phi_t \}_{t=1}^{t=t_{stop}}, \boldsymbol{Z}_{t_{stop}}} \mathcal{C}_{model}+\mathcal{C}_{para}+\mathcal{C}_{mal} \\
		s.t.\max\big\{\boldsymbol\varphi(t_{stop}) \big\}>\eta _{th},\label{eq20}
	\end{aligned}
\end{equation}
where the $\boldsymbol{Z}_{t_{stop}}$ is the set of clients who will be performed parameter audit.

\section{Deep Reinforce Learning Based ASS}
In the model audit phase, the actual status of the client (honest/malicious) is unknown to the server, and the server obtains noisy results by executing the model-audit process on some clients in each audit round. In this process, the actual state of the client has only partial observability. With this regard, we can map the selection audit process to partially observable Markov decision processes (POMDP). Moreover, a POMDP is usually PSPACE (Polynomial Space)-hard problems \cite{10.5555/2875343.2875347}. In general, the solution may require extensive computing resources and time, especially when facing complex practical problems. Deep reinforcement learning (DRL) has excellent advantages in solving these problems \cite{pmlr-v80-igl18a}.

In this section, we present  POMDP modeling on the client selection problem and propose a deep reinforcement learning based AI-enabled audit selection strategy (DRL-ASS) to select clients for the Model-Audit in each round, thereby minimizing system overhead and ensuring audit accuracy.
\subsection{Partially Observed Markov Decision Process}
POMDP is a mathematical framework   that combines the concepts of Markov decision process (MDP) and partial observability to handle the situations where partial uncertainty and observation limitations exist in the decision-making process. POMDP is usually described as a six-tuple ($\boldsymbol{S},\boldsymbol{A},\boldsymbol\Omega,\boldsymbol{\mathcal{O}},P,R$), where $\boldsymbol{S}$ denotes the set of states, $\boldsymbol{A}$ denotes the set of actions, and $\boldsymbol\Omega$ denotes the observation space, which includes all the possible observed values that agent can obtain. There is an implicit relationship between observations and system states in POMDP. $\boldsymbol{\mathcal{O}}$ is a conditional observation model that describes the probability of generating each possible observation given the system’s state. $P$ is a set of conditional transition probabilities between states. $R$ is the reward function, which indicates the reward obtained by taking an action in a specific state. In POMDP, the agent aims to choose the optimal action based on current observations and historical information. Since the system state is not fully observable, the agent must infer the system state from the observed values and use this information to choose the optimal action.
\paragraph{State}In our system, state $\boldsymbol{\mu}(t)\in\left\{0,1\right\}^N$ represents the identity information of $N$ clients at audit round $t$, where 0 and 1 mean honest and malicious, respectively. However, this state cannot be observed by the agent directly. 

\paragraph{Action space}Action space $\boldsymbol{A}$ is defined as the set of all possible decisions that the agent can make. In our system, an efficient action $a_t\in \boldsymbol{A}$  refers to selecting a subset of $N$ clients at audit round $t$ to perform the model audit algorithm on their uploaded local models. Specifically, the action space is an integer space with values ranging from 0 to $2^{N-1}$, and the N-bit binary of each action a corresponds to an audit selection, eg. In the scenario of $N=5$, $a_t=0$ corresponds to $\left\{0,0,0,0,0\right\}$, indicating that no client will be audited at audit round $t$, and $a_t=3$ corresponds to selecting $\left\{0,0,0,1,1\right\}$, indicating auditing clients 4 and 5 at audit round $t$. 

\paragraph{Observation space}The observation space $\boldsymbol\Omega$ is defined as the set of all the possible observed information. This information is usually related to the system state $\boldsymbol{\mu}(t)$. In our system, the observation state $\boldsymbol\lambda(t)\in\left\{0,1\right\}^{|Bin_N(a_t)|}$ is represented as the audit result obtained after executing the model audit on the selected set of clients, $|Bin_N(a_t)|$ represents the total number of clients selected in the $t$-th audit round. 

\paragraph{Conditional observation model}Conditional observation model $\mathcal{O}_t\in \boldsymbol{\mathcal{O}}$ can be modeled by the posterior belief vector $\boldsymbol{\varphi}(t)$.

\paragraph{Conditional Transition Probability}The condition transition probability is given by
\begin{equation}
	\begin{aligned}
		\mathbb{P}(\mathcal{O}_{t+1} | \mathcal{O}_t ,a_t)
		&=\mathbb{P} \big\{\boldsymbol\varphi(t+1)\big| \boldsymbol\varphi(t) ,\boldsymbol\lambda _{\Phi _t} (t)\big\} \\
		&=\prod_{u=0}^{h-1} \frac{\varphi_u(t)*\vartheta _u(t)}{\sum_{k=0}^{h-1}\varphi_k(t)*\vartheta_k(t)},\label{eq21}
	\end{aligned}
\end{equation}

\paragraph{Reward Function}Considering our design goal, we need to formulate an appropriate reward function to guide the agent in choosing actions. Our goal is to ensure the model audit accuracy while reducing system overhead to the greatest extent. Meanwhile, we notice that the actual client state is unknowable, and the noise in the audit process may lead the agent to accept wrong assumptions. Therefore, the reward should be set to achieve two goals: exploratory to find the true hypothesis and reduce total system overhead. Taking inspiration from the Average Bayesian Log-Likelihood Ratio (ABLLR) introduced in \cite{WOS:000461021200104}, we define the reward function as follows: 
\begin{equation}
	\begin{aligned}
{{r}_{t}}&=\xi (\underset{i=0}{\overset{h-1}{\mathop \sum }}\,{{\varphi }_{i}}\left( t \right)log\frac{{{\varphi }_{i}}\left( t \right)}{1-{{\varphi }_{i}}\left( t \right)}\\
&-\underset{i=0}{\overset{h-1}{\mathop \sum }}\,{{\varphi }_{i}}\left( t-1 \right)log\frac{{{\varphi }_{i}}\left( t-1 \right)}{1-{{\varphi }_{i}}\left( t-1 \right)})\\
&-\left( 1-\xi  \right)\left| Bi{{n}_{N}}\left( {{a}_{t}} \right) \right|.\label{eq22}
	\end{aligned}
\end{equation}
The first part quantifies the difference between the ABLLR functions in the two states. Concurrently, the second segment $\left| Bi{{n}_{N}}\left( {{a}_{t}} \right) \right|~$represents the count of clients engaged in model audit within the current state. By introducing the regularization parameter $\xi$, this reward function guides the agent to infer an audit selection policy with low overhead. 

\paragraph{Initialization and Termination}In our system, the initial state ${{\mathcal{O}}_{0}}$ can be represented by the prior probability that the server is malicious to the client. The termination state is the observation model corresponding to ${{\mathcal{O}}_{{{t}_{stop}}}}$. At the same time, to facilitate network training, we introduce the maximum number of audit round $L$; that is, when the number of audit round reaches$~L$, sampling will be terminated regardless of whether the $\max \left\{ \boldsymbol\varphi \left( L \right) \right\}>{{\eta }_{th}}$.

Under the given initial observation model ${{\mathcal{O}}_{0}}$, the agent chooses to perform an action ${{a}_{t}}\text{ }\!\!~\!\!\text{ }$according to the policy ${{\pi }_{\theta }}$ and obtains the reward ${{r}_{t}}$, and updates ${{\mathcal{O}}_{t}}\text{ }\!\!~\!\!\text{ }$to ${{\mathcal{O}}_{t+1}}$ until the terminal state is reached. In our system, the goal is to maximize the cumulative reward,
\begin{equation}
	\begin{aligned}
	R=\mathbb{E}\left[ \underset{t=0}{\overset{t=L}{\mathop \sum }}\,{{\gamma }^{t}}{{r}_{t}} \right], \label{eq23}
	\end{aligned}
\end{equation}
where $\gamma$ is the discount factor.

However, the client selection problem is influenced by the distribution of malicious clients and the intricate dynamics of the environment, such as the constraints imposed by various overhead parameters and the statistical randomness stemmed from audit error. Additionally, within the framework of POMDP, the state of the environment remains partially unobservable and the action search space corresponding to the problem scales exponentially with the number of clients involved. These factors pose a substantial challenge for algorithms based deep reinforcement learning (DRL), making it difficult to converge toward an optimal solution.

To effectively address this challenge and derive an audit selection strategy, the critical task of selecting a suitable DRL algorithm assumes paramount importance. 

\subsection{Diffusion Model}
The diffusion model \cite{ho2020denoising}, categorized as a deep generative model, has gained widespread utilization in contemporary years within the realm of image generation \cite{croitoru2023diffusion}. Its idea is to systematically perturb the distribution in the image data through the forward diffusion process and then learn the distribution of the perturbation added by the forward process through the reverse diffusion process to restore the original data distribution. 

The effect of the forward process is to perturb the data. It gradually adds Gaussian noise to the original data according to the pre-designed noise distribution until the data distribution tends to the standard Gaussian distribution. Let ${{x}_{0}}\sim g(x_0)$  denote the original data and its associated probability distribution. The forward process can then be expressed as follows:
\begin{equation}
	\begin{aligned}
   g\left( {{x}_{1}},{{x}_{2}},\ldots ,{{x}_{Y}}\text{ }\!\!|\!\!\text{ }{{x}_{0}} \right)=\underset{y=1}{\overset{Y}{\mathop \prod }}\,g\left( {{x}_{y}}|{{x}_{y-1}} \right),\label{eq24} 
	\end{aligned}
\end{equation}
\begin{equation}
	\begin{aligned}
		g\left( {{x}_{y}}\text{ }\!\!|\!\!\text{ }{{x}_{y-1}} \right)=\mathcal{N}\left( {{x}_{y}};\sqrt{1-{{\beta }_{y}}}{{x}_{-1}},{{\beta }_{y}}\mathbf{I} \right),\label{eq25} 
	\end{aligned}
\end{equation}
where ${{x}_{y}}$ denotes the sample after adding perturbation at diffusion step $y$ and $Y$ is the max diffusion step. And ${{\beta }_{y}}$ is the given variance parameter. 
Reverse process is defined as a Markov chain with learned Gaussian transitions starting at ${{p}_{\theta }}\left( {{x}_{Y}} \right)=~\mathcal{N}\left({{x}_{Y}};0,\mathbf{I} \right)$:
\begin{equation}
	\begin{aligned}
	{{p}_{\theta }}\left( {{x}_{Y:0}} \right)={{p}_{\theta }}\left( {{x}_{Y}} \right)\underset{y=1}{\overset{Y}{\mathop \prod }}\,{{p}_{\theta }}\left( {{x}_{y-1}}\text{ }\!\!|\!\!\text{ }{{x}_{y}} \right),\label{eq26} 
	\end{aligned}
\end{equation}
\begin{equation}
	\begin{aligned}
	 {{p}_{\theta }}\left( {{x}_{y-1}}\text{ }\!\!|\!\!\text{ }{{x}_{y}} \right)=N\left( {{x}_{y-1}};{{\mu }_{\theta }}\left( {{x}_{y}},y \right),{{\text{ }\!\!\Sigma\!\!\text{ }}_{\theta }}\left( {{x}_{y}},y \right) \right),\label{eq27} 
	\end{aligned}
\end{equation}
where ${{\mu }_{\theta }}\left( {{x}_{y}},y \right)$ and ${{\text{ }\!\!\Sigma}_{\theta }}\left( {{x}_{y}},y \right)$ denote the mean and variance of Gaussian distribution, respectively. Through iterative learning, a distribution close to the original data distribution ${{x}_{0}}~$can be obtained.

\subsection{AI-Enabled Audit Selection strategy (ASS)}
Next let us elaborate on how the based diffusion model AI-enabled audit selection strategy (ASS) generates a selection audit policy in the HiAudit-FL system. Specifically, we aim to train a diffusion model-based optimizer to generate optimal discrete decision solutions. The decision-making solution can be expressed as a set of discrete probabilities for selecting each action in the action space under different states $\mathcal{O}$.

Combining the forward process and reverse process of the diffusion model, in our system, the training goal is to obtain the optimal selection decision ${\boldsymbol{x}_{0}}={{\pi }_{\boldsymbol\theta }}\left( \mathcal{O} \right)$, which outputs the probability distribution of each act being selected in state $\mathcal{O}$.

\paragraph{Forward process} Based on a target probabilities policy ${\boldsymbol{x}_{0}}$, a sequence of Gaussian perturbations is added at each diffusion step $y\left( 0\le y\le Y \right)$ and obtain$\text{ }\!\!~\!\!\text{ }{\boldsymbol{x}_{1}},{\boldsymbol{x}_{2}},\ldots ,{\boldsymbol{x}_{Y}}$. In the process, the transition from  ${\boldsymbol{x}_{y-1}}$ to ${\boldsymbol{x}_{y}}$ can be defined as a Gaussian distribution $g\left( {\boldsymbol{x}_{y}}\text{ }\!\!|\!\!\text{ }{\boldsymbol{x}_{y-1}} \right)$with mean and variance $\sqrt{1-{{\beta }_{y}}}{\boldsymbol{x}_{y-1}}$ and ${{\beta }_{y}}\mathbf{I}$, respectively. Although the forward process is not executed actually in our system, it helps reveal the mathematic relationship between target policy  ${\boldsymbol{x}_{0}}~$ and any  ${\boldsymbol{x}_{y}}$ as follows, 
\begin{equation}
	\begin{aligned}
	  {\boldsymbol{x}_{y}}=\sqrt{{{{\bar{\omega}}}_{y}}}{\boldsymbol{x}_{0}}+\sqrt{1-{{{\bar{\omega }}}_{y}}}\boldsymbol\epsilon ,\left( \forall y=1,2,\ldots Y \right),\label{eq28} 
	\end{aligned}
\end{equation}

\paragraph{Reverse process} In our system, the optimal audit selection policy training mainly relies on the reverse process, the purpose is to infer the distribution policy corresponding to the optimal audit selection policy from the perturbation sample ${\boldsymbol{x}_{y}}\sim\mathcal{N}\left( \boldsymbol{0},\mathbf{I} \right)$. The reverse transition ${{p}_{\theta }}\left( {\boldsymbol{x}_{y-1}}\text{ }\!\!|\!\!\text{ }{\boldsymbol{x}_{y}} \right)$ from ${\boldsymbol{x}_{y}}$ to ${\boldsymbol{x}_{y-1}}$ as given by,
\begin{equation}
	\begin{aligned}
		 {{p}_{\boldsymbol\theta }}\left( {{x}_{y-1}}\text{ }\!\!|\!\!\text{ }{\boldsymbol{x}_{y}} \right)&=
		\mathcal{N}\left( {\boldsymbol{x}_{y-1}};{{\mu }_{\boldsymbol\theta }}\left( {\boldsymbol{x}_{y}},y,\mathcal{O} \right),{{{\tilde{\beta }}}_{y}}\mathbf{I} \right),\\&\left( \forall y=1,2,\ldots Y \right),\label{eq29} 
	\end{aligned}
\end{equation}
where
\begin{equation}
	\begin{aligned}
{{\mu }_{\boldsymbol\theta }}\left( {\boldsymbol{x}_{y}},y,\mathcal{O} \right)=&\frac{\sqrt{{{{\bar{\omega }}}_{y-1}}}{{\beta }_{y}}}{1-~{{{\bar{\omega }}}_{y}}}{\boldsymbol{x}_{0}}+\frac{\sqrt{{{{\bar{\omega }}}_{y}}}\left( 1-{{{\bar{\omega }}}_{y-1}} \right)}{1-~{{{\bar{\omega }}}_{y}}}{\boldsymbol{x}_{y}},  \\
&\left( \forall y=1,2,\ldots Y \right), \label{eq30} 
	\end{aligned}
\end{equation}
and
\begin{equation}
	\begin{aligned}
 {{{\tilde{\beta }}}_{y}}=\frac{1-{{{\bar{\omega }}}_{y-1}}}{1-{{{\bar{\omega }}}_{y}}}{{\beta }_{y}}, \label{eq31} 
	\end{aligned}
\end{equation}
where the model ${{\mu }_{\theta }}~$ is learned through deep networks, ${{\tilde{\beta }}_{y}}$ is a deterministic variance amplitude.

According to (28), target policy ${{x}_{0}}$ can be expressed as follows,
\begin{equation}
	\begin{aligned}
     {\boldsymbol{x}_{0}}=\frac{{\boldsymbol{x}_{y}}}{\sqrt{{{{\bar{\omega }}}_{y}}}}-&\sqrt{\frac{1}{{{{\bar{\omega }}}_{y}}}-1}*\tanh \left( {{\boldsymbol\epsilon }_{\boldsymbol\theta }}\left( {\boldsymbol{x}_{y}},y,\mathcal{O} \right) \right),
     \\&\left( \forall y=1,2,\ldots Y \right), \label{eq32} 
	\end{aligned}
\end{equation}
where ${{\boldsymbol\epsilon }_{\boldsymbol\theta }}\left( {{x}_{y}},y,\mathcal{O} \right)$ is a deep model parameterized by $\boldsymbol\theta$, which generates noise parameters that assist denoising under state $\mathcal{O}$. By substituting (28) into (30), the estimated mean of the reverse process can be obtained as follows,
\begin{equation}
	\begin{aligned}
{{\mu }_{\boldsymbol\theta }}\left( {\boldsymbol{x}_{y}},y,\mathcal{O} \right)=&\frac{1}{\sqrt{{{\omega }_{y}}}}\left( {\boldsymbol{x}_{y}}-\frac{{{\beta }_{y}}\tanh \left( {{\boldsymbol\epsilon }_{\boldsymbol\theta }}\left( {\boldsymbol{x}_{y}},t,\mathcal{O} \right) \right)}{\sqrt{1-~{{{\bar{\omega }}}_{y}}}} \right),  \\
&\left( \forall y=1,2,\ldots Y \right). \label{eq33} 
	\end{aligned}
\end{equation}
Finally, the target generation distribution ${{p}_{\boldsymbol\theta }}\left( {\boldsymbol{x}_{0}} \right)$ can be further obtained through (26).

To solve the problem that gradients cannot be backpropagated normally due to sampling from random distributions, we adopt a reparameterization method \cite{du2023generative} to decouple randomness from distribution parameters. Specifically, we use the following update rule instead,
\begin{equation}
	\begin{aligned}
	 {\boldsymbol{x}_{y-1}}={{\mu }_{\boldsymbol\theta }}\left( {\boldsymbol{x}_{y}},y,\mathcal{O} \right)+{{\left( \frac{{{{\tilde{\beta }}}_{y}}}{2} \right)}^{2}}\odot \boldsymbol\epsilon ,\left( \forall y=1,2,\ldots Y \right), \label{eq34} 
	\end{aligned}
\end{equation}
where$\text{ }\!\!~\!\!\text{ }\boldsymbol\epsilon \sim \mathcal{N}\left( \boldsymbol{0},\boldsymbol{I} \right)$.  By applying the backward update rule iteratively, we can finally get ${{x}_{0}}$ for the output. Next, we apply the softmax function to ${{x}_{0}}$ and convert it into the probability distribution ${{\pi }_{\boldsymbol\theta }}\left( \mathcal{O} \right)$:
\begin{equation}
	\begin{aligned}
		 {{\pi }_{\boldsymbol\theta }}\left( \mathcal{O} \right)=\left\{ \frac{{{e}^{x_{0}^{a}}}}{\mathop{\sum }_{k\in \boldsymbol{A}}{{e}^{x_{0}^{k}}}} \right\},\forall a\in \boldsymbol{A}. \label{eq35} 
	\end{aligned}
\end{equation}
To effectively train ${{\pi }_{\boldsymbol\theta }}\left( \mathcal{O} \right)$, we integrate the diffusion model-based ASS into the actor-critic architecture-based DRL framework. Specifically, under the critic network for guidance, the reverse-diffusion chain of ASS as the actor network to generate a higher-value policy distribution. Then, the ASS network and critic network are trained jointly.
\subsection{The DRL-ASS algorithm framework}
The framework of the proposed DRL-ASS algorithm is shown in Fig. 3., which includes five parts: environment, experience reply memory, ASS network, Critic network, and Network update process.
\begin{figure}[htbp]
	\centerline{\includegraphics[width=1\linewidth]{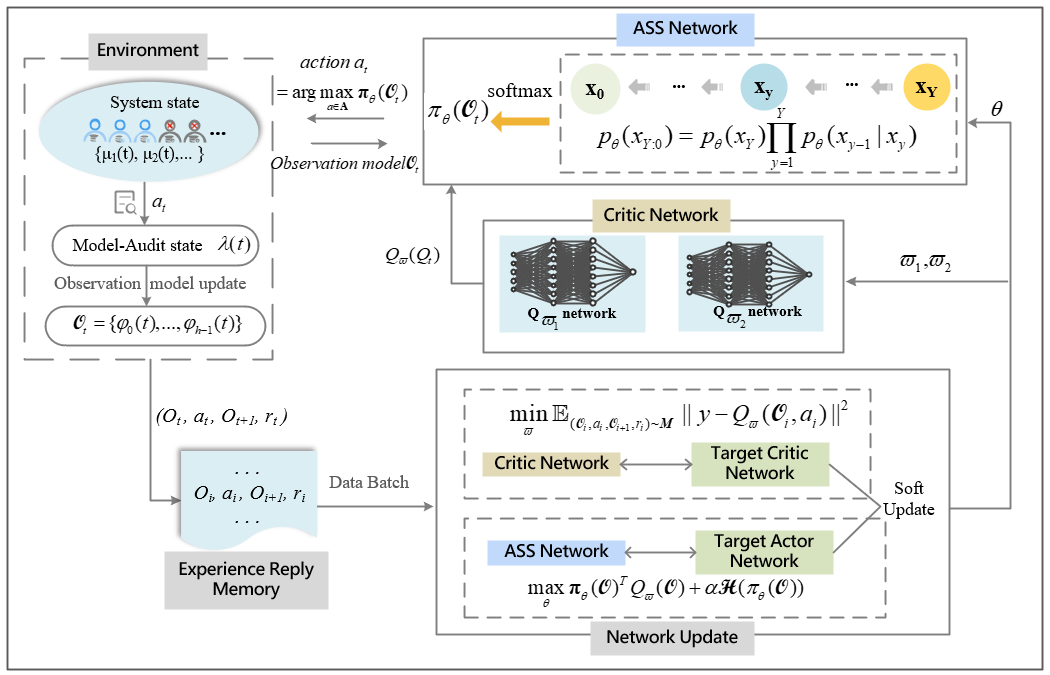}}
	\caption{DRL-ASS algorithm framework}
	\label{fig3}
\end{figure}
Without prior system state, given the initial observation model ${{\mathcal{O}}_{0}}$, the agent chooses to perform an action ${{a}_{t}}~$according to the policy ${{\pi }_{\boldsymbol\theta }}$ and obtains the model-audit results. Then, the agent gets an immediate reward ${{r}_{t}}$ and updates ${{\mathcal{O}}_{t}}$ to ${{\mathcal{O}}_{t+1}}$. Moreover, to effectively learn from historical experience, the agent stores the four-tuple$\left( {{\mathcal{O}}_{t}},{{a}_{t}},{{\mathcal{O}}_{t+1}},{{r}_{t}} \right)$ sampled from the environment as an experience each round $t$ in the reply memory buffer.

$\mathit{ASS~Network.}$ The ASS network functions as an actor network in the DRL paradigm. In each audit round $t$, it takes the observation model ${{\mathcal{O}}_{t}}$ as input and captures the dependency ${{\pi }_{\mathbf{\boldsymbol\theta }}}\left( \boldsymbol{A}|{{\mathcal{O}}_{t}} \right)~$between the observation model ${{\mathcal{O}}_{t}}$ and the action space $\boldsymbol{A}$ through the generation process of the diffusion model. Thus, the target policy is obtained.	
  
$\mathit{Critic~Network.}$ During the policy update process, the ASS network is trained and optimized by randomly sampling small batches of experience with size $M$ from the experience replay memory. Moreover, we use the Double Critic network as the policy evaluation function to reduce the overestimation bias. This Q function outputs a vector containing the Q values of all possible actions ${{a}_{t}}\in \boldsymbol{A}$, that is
\begin{equation}
	\begin{aligned}
	 {{Q}_{\boldsymbol\varpi }}\left( {{\mathcal{O}}_{t}} \right)=\min \left\{ ~{{Q}_{{{\boldsymbol\varpi }_{1}}}}\left( {{\mathcal{O}}_{t}} \right){{Q}_{{{\boldsymbol\varpi }_{2}}}}\left( {{\mathcal{O}}_{t}} \right) \right\}, \label{eq36} 
	\end{aligned}
\end{equation}
where ${\boldsymbol{\varpi }_{1}}$ and ${\boldsymbol{\varpi }_{2}}$ are the parameters of two critic networks, respectively, and these two networks are updated independently with the same optimization goal.

$\mathit{Policy~Network~Update.}$ To prevent the policy from overdetermining certain actions and prematurely converging to suboptimal solutions, an action entropy regularization term is introduced to encourage exploration. The training objectives are as follows,
\begin{equation}
	\begin{aligned}
\underset{\mathbf{\boldsymbol\theta }}{\mathop{\max }}\,{{\pi }_{\mathbf{\boldsymbol\theta }}}{{\left( {{\mathcal{O}}_{t}} \right)}^{T}}{{Q}_{\boldsymbol\varpi }}\left( {{\mathcal{O}}_{t}} \right)+\alpha H\left( {{\pi }_{\mathbf{\boldsymbol\theta }}}\left( {{\mathcal{O}}_{t}} \right) \right), \label{eq37} 
	\end{aligned}
\end{equation}
where $\mathcal{H}\left( {{\pi }_{\boldsymbol\theta }}\left( {{\mathcal{O}}_{t}} \right) \right)$ is the entropy of the action probability distribution ${{\pi }_{\boldsymbol\theta }}\left( {{\mathcal{O}}_{t}} \right)$, and the temperature coefficient $\alpha ~$controls the degree of exploration. 

$\mathit{Critic~Network~Update.}$ Effective training of the Q function can guide the agent to find the optimal strategy  . Therefore, we improve the Q function evaluation ability by minimizing the TD error between the Q-target and Q-value,
\begin{equation}
	\begin{aligned}
	\underset{\boldsymbol\varpi }{\mathop{\min }}\,{{\mathbb{E}}_{\left( {{\mathcal{O}}_{t}},{{a}_{t}},{{\mathcal{O}}_{t+1}},{{r}_{t}} \right)\sim M}}{{\left| \left| \hat{y}-{{Q}_{\boldsymbol\varpi }}\left( {{\mathcal{O}}_{t}},{{a}_{t}} \right) \right| \right|}^{2}}, \label{eq38} 
	\end{aligned}
\end{equation}
where,
\begin{equation}
	\begin{aligned}
	\hat{y}={{r}_{t}}+\gamma {{{\hat{\pi }}}_{\boldsymbol{\hat{\theta }}}}{{\left( {{\mathcal{O}}_{t+1}} \right)}^{T}}{{{\hat{Q}}}_{{\hat{\boldsymbol\varpi }}}}\left( {{\mathcal{O}}_{t+1}} \right), \label{eq39} 
	\end{aligned}
\end{equation}
where ${{\hat{\pi }}_{\boldsymbol{\hat{\theta }}}}$ and ${{\hat{Q}}_{{\hat{\varpi }}}}$represent the target actor network and the target critic network respectively, and they have consistent network structures with corresponding actor network ${{\pi }_{\theta }}$ and critic network ${{Q}_{\varpi }}$. Moreover, the parameters $\boldsymbol{\hat{\theta }}$ and $\hat{\varpi }$ of the target network are processed using the following soft update rules.
\begin{equation}
	\begin{aligned}
	\boldsymbol{\hat{\theta }}\leftarrow \iota \boldsymbol\theta +\left( 1-\iota  \right)\boldsymbol{\hat{\theta }}, \label{eq40} 
	\end{aligned}
\end{equation}
\begin{equation}
	\begin{aligned}
		\hat{\boldsymbol\varpi }=\iota \hat{\boldsymbol\varpi }+\left( 1-\iota  \right)\hat{\boldsymbol\varpi }, \label{eq41} 
	\end{aligned}
\end{equation}
$\iota \in \left( 0,1 \right]$ is the hyperparameter for soft update.

We summarize the DRL-ASS algorithm process in Algorithm 1.
	\begin{algorithm}
	\begin{minipage}{\linewidth}
		\caption{DRL-ASS Algorithm for Clients Audit Selection}
		Initialize policy parameters $\boldsymbol\theta$, Q-function parameters $\boldsymbol\varpi$, target network parameters $\hat{\boldsymbol\theta } \gets \boldsymbol{\theta},\hat{\boldsymbol\varpi } \gets \boldsymbol\varpi$,  number of clients $N$;
		\begin{algorithmic}
			\item[]
			\For{Traning step \textbf{from} $1$ \textbf{to} MaxStep}{
				\item[]
				Initialize state  $\mathcal{O}$${_0}$;\\
				Generate a new hypothesis $H_{true}$ to be true according to the prior belief $\mathcal{O}_0$;\\
				$t=0$;
				\While{$\mathcal{O}_t$ is not a terminal state}{
					\item[]
					Observe state $\mathcal{O}$${_t}$ and initialize a random normal distribution $\mathbf{x}_Y\sim\mathcal{N}(0, \mathbf{I})$;
					\For{the diffudion step $y$ \textbf{from} $Y$ \textbf{to} $1$}{
						\item[]
						Infer and scale a denoising distribution $\tanh(\boldsymbol\epsilon_{\boldsymbol\theta }(\mathbf{x}_y,y,\mathcal{O}_y))$ using a deep neural network;\\
						Calculate the mean $\mu_{\boldsymbol\theta}$ of the reverse transition distribution $p_{\boldsymbol\theta}(\mathbf{x}_{y-1}|\mathbf{x}_y)$;\\
						Calculate the distribution $\mathbf{x}_{y-1}$ using the reparameterization trick;
					}
					\item[]
					Calculate the probability distribution of $\mathbf{x}_0$ and select action $a_t$ at random based on it;\\
					Execute action $a_t$ and receive the audit results  $\lambda(t)$, then update the next observation model $\mathcal{O}_{t+1}$ and obtain reward $r_t$;\\
					Store the tuple \{$\mathcal{O}_{t}$, $a$$_t$,$r{_t}$ in the replay memory buffer.
				}
				\item[]
				Sample a batch of transitions $\mathbf{M}  = {(\mathcal{O}_i, a_i, \mathcal{O}_{i+1}, r_i)}$ from the replay buffer;\\
				Update the parameters $\boldsymbol\theta$, $\hat{\boldsymbol\theta}$,$\boldsymbol\varpi$, $\hat{\boldsymbol\varpi}$;
			}
			\item[]
			Save the trained neural networks parameter $\boldsymbol\theta^\ast$ and $\boldsymbol\varpi^\ast$.
		\end{algorithmic}
	\end{minipage}
\end{algorithm}

\section{Numerical Results and Discussions}
In this section, we use simulation experiments  to evaluate the performance of the proposed Hi-Audit FL system and the client selection audit algorithm based on DRL-ASS. 
\subsection{Simulation  Settings}
\subsubsection*{\bf Experimental Platform} We set up the simulation experiment on a platform with an Intel i512600KF CPU with 48 GB RAM. The software environment used is Python3.8, GPU NVIDIA GeForce RTX 3070, and frameworks such as PyTorch are used to simulate the proposed HiAudit-FL system and the client selection audit algorithm based on DRL-ASS.

\subsubsection*{\bf Scenario} We simulate a scenario with  $N=5$ clients, and each client has a dataset with sample size 100. We aim to train an agent to select several clients to perform the model audit in each audit round. There are $h=32\text{ }\!\!~\!\!\text{ }$hypotheses in the agent. We assume that the agent can know no prior information about the malicious clients before the audit process starts. Thus, the initial observation model$~~{{\mathcal{O}}_{0}}\text{ }\!\!~\!\!\text{ }$is set as [$\frac{1}{h},\frac{1}{h},\ldots ,\frac{1}{h}$]. The number of CPU cycles required for performing parameter audit on one sample is randomly chosen within $\left[ 1,4 \right]\times {{10}^{4}}$ cycles/sample \cite{9798758}. And the number of CPU cycles required for performing model audit on one local model is randomly chosen within $\left[ 1,2 \right]\times {{10}^{2}}$ cycles/model. The regularization parameters $\xi =0.4$, and the probability of error in the model audit method  .

\subsubsection*{\bf Model Design} The DRL-ASS adopts the diffusion model-based ASS network as the core of the actor network and uses two critic networks with the same structure to alleviate the overestimation problem. The actor network consists of 3 fully connected layers using the Mish activation function and a Tanh activation function to normalize the output. The number of neurons in the fully connected layer is 32, 256, and 256, respectively. Its output is the probability of all selection actions in a given state. The diffusion step of the diffusion model is set to$\text{ }\!\!~\!\!\text{ }Y=5$, and the learning rate of the actor network is $1\times {{10}^{-4}}$. The critic network consists of two fully connected layers activated by the Mish function. Both of the neurons are set to 256. The learning rate of the critic network is set to $1\times {{10}^{-3}}$. For the hyperparameters used in the training process, we set the action entropy regularization parameter $\alpha =0.05$ and the discount rate $\gamma =0.95$.

\subsubsection*{\bf Benchmarks} We compare the proposed selection strategy with two DRL algorithms, deep recurrent Q-Network (DRQN) \cite{hausknecht2015deep} and soft actor-critic (SAC) \cite{haarnoja2018soft} as well as a heuristic algorithm, Random policy. DRQN is a classic algorithm for solving the POMDP problem, while SAC is efficient in handling high-dimensional state space and action space environments. Random Policy performs the audit process by randomly selecting some clients in each audit round.

In assessing the performance of our algorithm, we focus on two key metrics: the misjudgment rate and computing overhead. The misjudgment rate is a crucial indicator of the audit accuracy within our multi-round audit mechanism. It is defined by the following formula:
\begin{equation}
	\begin{aligned}
	misjudgment~rate=\frac{{{c}_{err}}}{100}, \label{eq42} 
	\end{aligned}
\end{equation}
where ${{c}_{err}}$ represents the aggregate number of errors identified throughout a series of 100 audit tests. Errors are recorded in two specific instances: firstly, when an honest client is incorrectly classified as malicious at the end of the audit process, and secondly, when any malicious clients remain within the system at the end of the audit.$~$The overhead reflects the total computing overhead during the execution of the hierarchical audit mechanism. An efficient client selection strategy is vital in maintaining low computing overhead while simultaneously keeping the misjudgment rate at a low level. These two metrics collectively provide a comprehensive evaluation of the client selection strategy's rationality and efficacy.

\subsection{Numerical Results and  Analysis}
First, we compare  the reward of different algorithms during the training phase. Fig. 4. shows the reward curves of the ASS strategy and three baseline strategies in the training phase. Compared to the Random strategy, we find that DRL-ASS, SAC, and DRQN can adapt to the changing environment and capture the relationship between action selection and utility in different states to increase the reward. The SAC-based strategy and the DRQN-based strategy can reach a stable value in fewer training rounds, but they are unable to continuously improve the reward  by adjusting the strategy. On the contrary, DRL-ASS can fine-tune the model output in multiple steps by performing a reverse process in the ASS network, thereby improving sampling quality. Therefore, DRL-ASS outperforms  the other three baseline algorithms.
\begin{figure}[htbp]
	\centerline{\includegraphics[width=1\linewidth]{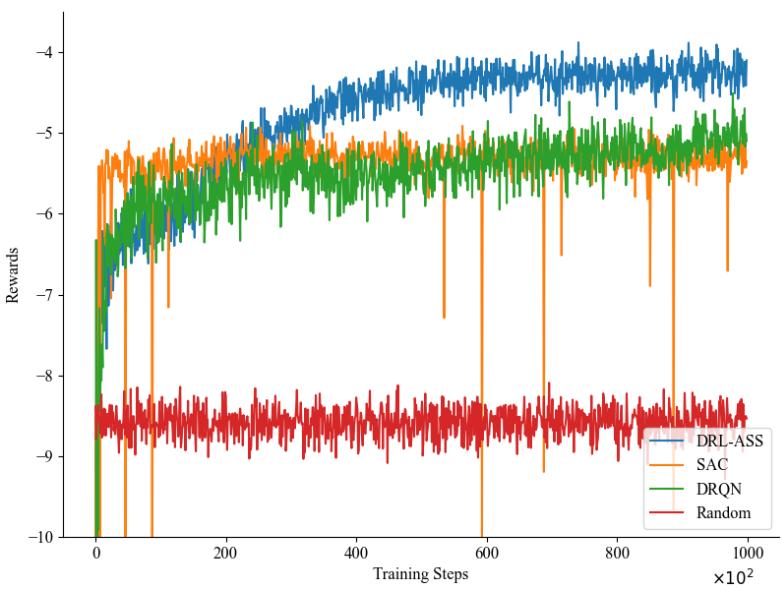}}
	\caption{The training process of different algorithm, with ${{\eta }_{th}}=0.8$.}
	\label{fig4}
\end{figure}
\begin{figure}[htbp]
	\centerline{\includegraphics[width=1\linewidth]{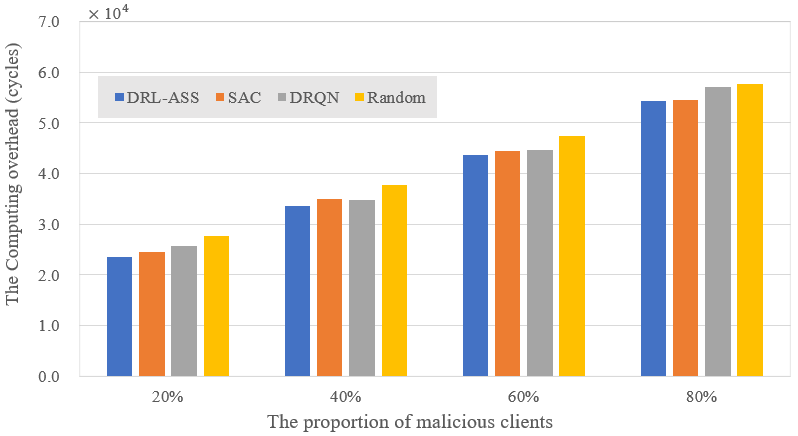}}
	\caption{System overhead of different algorithms}
	\label{fig5}
\end{figure}
\begin{table}[ht]
	\caption{Algorithm Performance with Varying Malicious Client Ratios}
	\label{tab:algorithm_performance}
	\centering
	\begin{tabular}{|c|c|c|c|c|}
		\hline
		Mal. Client \% & DRL-ASS & SAC & DRQN & Random \\ \hline
		20\% & 0.05 & 0.13 & 0.12 & 0.11 \\ 
		40\% & 0.08 & 0.09 & 0.13 & 0.13 \\ 
		60\% & 0.09 & 0.09 & 0.11 & 0.12 \\ 
		80\% & 0.06 & 0.13 & 0.06 & 0.12 \\ \hline
	\end{tabular}
\end{table}
Next, we compare the performance of the strategies during the test process with different proportions of malicious clients in the environment (20\%, 40\%, 60\%, 80\%), We test each strategy 100 times for each case and record the average total overhead and average audit misjudgments. Here the misjudgment rate refers to the proportion of the number of times all malicious clients are audited to the total number of tests when the model audit threshold ${{\eta }_{th}}$ is reached. The simulation results are shown in Fig. 5. and Table I. We can see that the SAC strategy maintains a low overhead level in all cases, but for the proportion of malicious clients is 20\% and 80\%, we observe a notable increase in the audit misjudgment rate. By observing the actions of the SAC policy we can find that it keeps selecting the same set of clients to perform audits in each round. This repetitive selection fails to effectively adapt to varying states, hindering its ability to accurately discern the dynamic relationship between state and action selection. Conversely, the DRQN-based strategy tends to select and audit a small number of nodes each time. While this approach minimizes the immediate overhead, it inadvertently leads to inaccuracies in updating confidence information. Consequently, this strategy not only increases the audit overhead but also elevates the audit misjudgment rate. The DRL-ASS strategy maintains a low misjudgment rate and low audit overhead in different cases. Its advantage can be attributed to its use of DRL-ASS based on the diffusion model, which enhances its ability to capture complex observation patterns. 

Then, we compare the impact of different blocking thresholds on the misjudgment rate and computing overhead. Fig. 6. shows that, in different proportions of malicious clients cases, the misjudgment rate roughly decreases with the blocking threshold. This trend indicates that the misjudgment rate can be modulated by the blocking threshold in a certain extent, aligning with the initial objective of our problem formulation. At the same time, the results in Fig. 7. show that as the confidence threshold increases, the total audit overhead will accordingly increase. This is because that the agent needs to collect more audit model samples during the audit process to increase the maximum confidence, thereby increasing the model audit overhead.
\begin{figure}[htbp]
	\centerline{\includegraphics[width=1\linewidth]{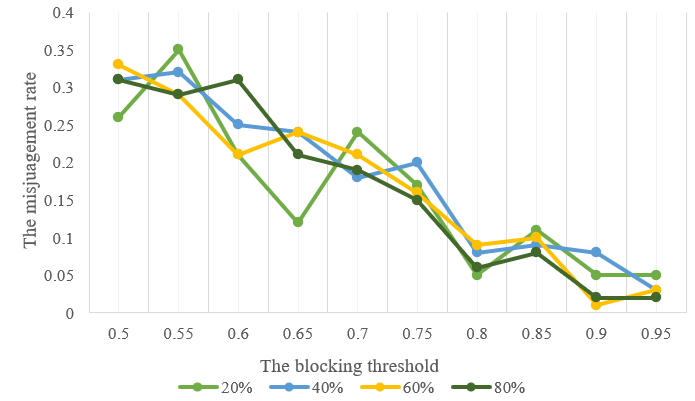}}
	\caption{The blocking threshold versus the misjugement rate for different malicious client proportions.}
	\label{fig6}
\end{figure}
\begin{figure}[htbp]
	\centerline{\includegraphics[width=1\linewidth]{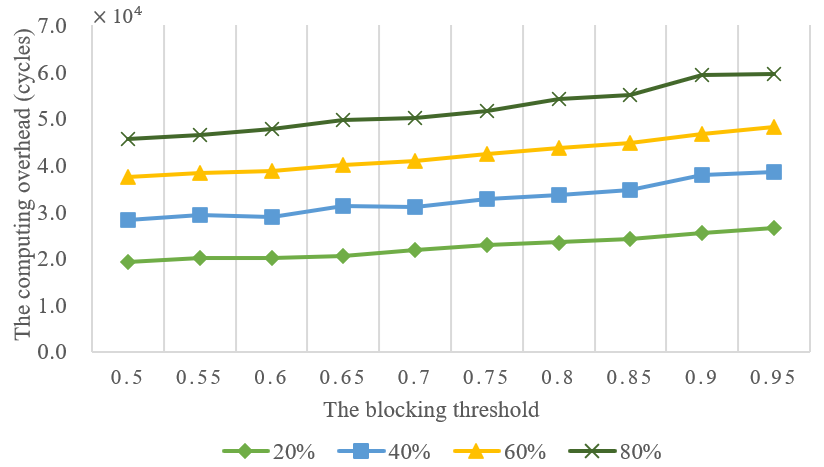}}
	\caption{The blocking threshold versus the compution overhead for different malicious client proportions.}
	\label{fig7}
\end{figure}

Finally, we compare the proposed HiAudit mechanism based on DRL-ASS with the audit methods of Only Parameter-Audit and Only Model-Audit. As illustrated in Fig. 8., we note that the misjudgment rate for the Only Model-Audit method stays within the range of 0.35$\sim$0.4 across various cases, which means that Only Model-Audit cannot distinguish malicious clients and take timely measures to reduce their impact on the learning process. Conversely, for different numbers of malicious clients, the HiAudit mechanism based on DRL-ASS maintains the misjudgment rates below 0.1. This underlines the mechanism's efficacy in accurately distinguishing between malicious and honest clients. As for the Only Parameter-Audit method, it shows a low misjudgment rate in different cases. However, this method results in considerable overhead due to substantial parameter retraining and complex auditing for each client. This makes it less efficient compared to other methods. 

As depicted in Fig. 9., the proposed HiAudit mechanism based on DRL-ASS exhibits adaptability to changes in cases. In our specific cases, compared with Only Model-Audit method, the overhead of the HiAudit mechanism based on DRL-ASS is reduced by 6.1\%, 23.9\%, 32.3\% and 36.1\%, respectively. Notably, in the cases with a low proportion of malicious clients, the Only Model-Audit method exhibits a low overhead. However, due to the limitation of audit error limitations, it fails to timely identify and eliminate malicious clients, leading to a significant increase in retention overhead in the presence of multiple malicious entities. Besides, when the proportion of malicious clients does not exceed 60\%, the total overhead of the HiAudit mechanism based on DRL-ASS is lower than that of Only Parameter-Audit method. Specifically, in cases where the proportion of malicious clients is 20\%, 40\%, and 60\%, the total overhead reduction of the proposed HiAudit mechanism compared to the Only Parameter-Audit method is 53\%, 32.5\%, and 12\%, respectively. When the proportion of malicious is 80\%, Only Parameter-Audit holds a slight advantage, suggesting its suitability in scenarios with a large-scale presence of malicious clients. Nevertheless, as the server typically cannot ascertain the number of malicious clients in advance, the proposed HiAudit mechanism based DRL-ASS proves to be more effective in reducing the system's total overhead through its adaptive audit strategies. 
\begin{figure}[htbp]
	\centerline{\includegraphics[width=1\linewidth]{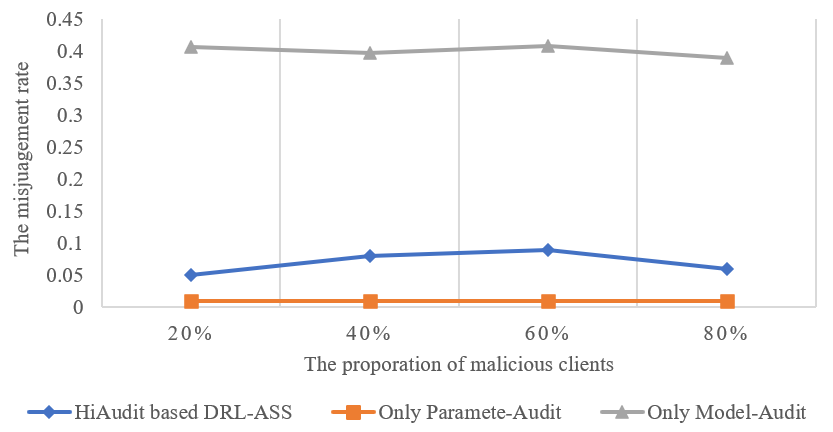}}
	\caption{Audit misjudgment rate of different audit mechanism.}
	\label{fig8}
\end{figure}
\begin{figure}[htbp]
	\centerline{\includegraphics[width=1\linewidth]{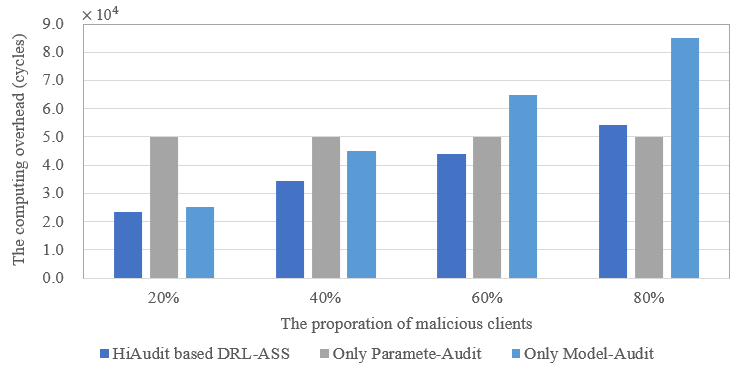}}
	\caption{System overhead of different audit mechanisms.}
	\label{fig9}
\end{figure}

\section{Conclusion}
In this work, we focused on enhancing the trustworthiness
of federated learning (FL) systems against malicious client
attacks with lower system overhead. To achieve this, we
integrated a novel audit mechanism called HiAudit, which
combines model audit method with parameter audit method,
into the existing FL framework to find out the malicious clients
accurately and efficiently. The primary challenge addressed in this study revolves around determining how to select clients for model audit and parameter audit, ensuring high audit accuracy while minimizing computational overhead. A
diffusion model-based Deep Reinforcement Learning Audit
Selection Strategy (DRL-ASS) is obtained to optimize the
model audit nodes selection process and effectively balance
system overhead with audit accuracy. Simulation results have
shown the feasibility of the proposed HiAudit mechanism
and the superiority of the proposed DRL-ASS algorithm.
The system overhead and audit accuracy could be effectively
balanced, which is a key issue in a trusted FL system.This
research not only presents a practical solution to enhancing the
security of FL systems but also contributes valuable insights
and methodologies that can be applied to similar problems
in the field. The significance of our work lies in its potential
to influence future developments in secure federated learning,
offering a robust framework that adeptly balances efficiency
and accuracy in a complex and dynamic environment.

\bibliographystyle{IEEEtran}
\bibliography{ref}

\end{document}